\documentclass[amsmath,amssymb,aps,a4paper,twocolumn]{revtex4-2}

\usepackage{jabbrv}
\usepackage[pdftex,colorlinks=true,bookmarks=false,citecolor=blue,urlcolor=blue]{hyperref} 

\usepackage{graphicx}
\usepackage{xcolor}
\usepackage[utf8]{inputenc}

\begin{document}

\title{Real-time operation of a multi-rate, multi-protocol quantum key distribution transmitter}

\author{Innocenzo De Marco,\textsuperscript{1,2,*} Robert I. Woodward,\textsuperscript{1} George L. Roberts,\textsuperscript{1,3} Taofiq K. Para\"{i}so,\textsuperscript{1} Thomas Roger,\textsuperscript{1} Mirko Sanzaro,\textsuperscript{1} Marco Lucamarini,\textsuperscript{1} Zhiliang Yuan,\textsuperscript{1} and Andrew J. Shields\textsuperscript{1}}

\affiliation{\textsuperscript{1}Toshiba Europe Ltd, 208 Cambridge Science Park, Cambridge, CB4 0GZ, United Kingdom\\
\textsuperscript{2}School of Electronic and Electrical Engineering, University of Leeds, Leeds, LS2 9JT, United Kingdom\\
\textsuperscript{3}Cambridge University Engineering Department, 9 JJ Thomson Avenue, Cambridge, CB3 0FA, United Kingdom}

\email{innocenzo.demarco@crl.toshiba.co.uk}

\begin{abstract}
Quantum key distribution (QKD) is the best candidate for securing communications against attackers, who may in the future exploit quantum-enhanced computational powers to break classical encryption.
As such, new challenges are arising from our need for large-scale deployment of QKD systems.
In a realistic scenario, transmitting and receiving devices from different vendors should be able to communicate with each other without the need for matching hardware.
Therefore, practical deployment of QKD would require hardware capable of adapting to different protocols and clock rates.
Here, we address this challenge by presenting a multi-rate, multi-protocol QKD transmitter linked to a correspondingly adaptable QKD receiver.
The flexibility of the transmitter, achieved by optical injection locking, allows us to connect it with two receivers with inherently different clock rates.
Furthermore, we demonstrate the multi-protocol operation of our transmitter, communicating with receiving parties employing different decoding circuits.
\end{abstract}

\maketitle

\section{Introduction}

Quantum key distribution (QKD) allows users to communicate with information theoretical security~\cite{Scarani.2009}.
It has become a strong candidate to resolve the imminent threat posed by quantum computers\cite{Arute.2019} to many existing cryptographic protocols based on complexity theory~\cite{Shor.1999}.
QKD, on the other hand, bases its security on the laws of quantum mechanics and would not be affected by the advent of quantum computers.
There have been many impressive demonstrations of point-to-point QKD over fibre links, including key sharing at 10~Mbit/s~\cite{Yuan.2018} and at a distance of 421~km~\cite{Boaron.2018} for a point-to-point link of optical fibre.
Such distances can be further improved thanks to the novel twin-field QKD protocol~\cite{Lucamarini.2018}, which has the capability to reach more than 500 km~\cite{Minder.2019,Fang.2020}.

Large-scale implementations of QKD will likely see users having different devices from different manufacturers.
This calls for an urgent need for interoperability.
In a realistic scenario, users would likely choose their device based on required performance and system cost, where different vendors might offer devices operating at different clock rates or via different protocols.~\cite{Moseley.2004}
Much of the current research within QKD aims at improving specific systems, while little consideration is given to interoperability between different systems.
This has led to the situation where dedicated hardware is required to implement separate protocols and to operate at a fixed clock rate.
Multi-rate, multi-protocol capability of the transmitters and receivers are hence highly desirable in this scenario~\cite{Korzh.2013,Sibson.2017b}.
However, much of the research in this direction has focussed solely on highlighting the flexibility of the transmitter by implementing the protocols separately.
In a scenario where communication with several parties is required, being able to switch between clock rates and protocols in real time is crucial for efficient communication.

In this manuscript we demonstrate real-time, multi-clock and multi-protocol continuous operation of a directly phase-modulated QKD transmitter~\cite{Yuan.2016,Paraiso.2019}.
By changing only the driving signal sent to our transmitter,we are able to change its operating regime in real time to communicate with a different receiving device.
The system stabilisation happens within a few seconds, without loss of integrity or degradation of the Quantum Bit Error Rate (QBER) and Secure Key Rates (SKR) afterwards.

\section{Experimental Realisation}
\paragraph*{Protocols}
We implement three QKD protocols.
The first is the differential phase shift (DPS) protocol~\cite{Inoue.2003}.
This protocol is based on the encoding of information in the phase difference of consecutive pulses.
Alice can encode her information with $\{0,~\pi\}$ phase shifts, and Bob will decode it using an asymmetric Mach-Zehnder interferometer (aMZI).

The second protocol is the time-bin encoded BB84 protocol~\cite{Bennett.1984} with decoy states~\cite{Lo.2005,Hwang.2003,Wang.2005}.
Here, the information is carried by the phase difference in pulse pairs.
Pulses belonging to different pairs need to have a random phase difference.
Such randomness is necessary to minimise the amount of information an eavesdropper can retrieve from the pulse~\cite{Lo.2007}.
Alice encodes her qubits in two different bases: for the $\mathbf{X}$ basis, phase shifts of $\{0,~\pi\}$ are used.
For the $\mathbf{Y}$ basis, phase shifts of $\{\pi/2,~3\pi/2\}$ are used.
The intensity of the optical pulses is modulated to implement the decoy states technique.

Finally, the last protocol we implement is the Coherent One Way (COW) protocol~\cite{Stucki.2005}.
The information is encoded in the time bins of pulse pairs, the coherence between which is used to check whether these pulses have been tampered with.

As a proof-of-principle demonstration, we consider the protocols' asymptotic secure key rates.
A full finite-size analysis is outside the scope of the present paper and would not change the significance of our results.
Moreover, to simplify the data collection, data is collected in one basis only for the BB84 protocol, with the assumption that the two bases are chosen with identical probabilities and have similar phase errors.
Finally, since the data is recorded and measured as a cumulative histogram of a repeating pattern, one detector is sufficient to measure the QBER.
This all means that we only need one detector for the BB84 and DPS protocols and two detectors for the COW protocol (one for the time-bin encoding, one for decoy detection).
This also means that the same receiver can be used for all three protocols at a given clock rate.
The input of the chip is split between the straight waveguide and the interferometer, with a $\sim~50:50$ splitting ratio, which allows us to measure all the quantities we need at once.
While not the case in a realistic scenario, this is reasonable for a proof of principle experiment as we are able to extract all the needed information from this simplified setup.

It is important to note that security proofs for the three protocols show that they have significant differences in their security.
For example, the decoy-state BB84 protocol is secure against coherent attacks, whereas the security proofs for the DPS and COW protocols are based on collective attacks.
For this reason, the comparison between the SKRs obtained with the BB84~\cite{Ma.2005}, DPS~\cite{Waks.2006} and COW~\cite{Stucki.2009} protocols should be taking this into account.

\paragraph*{Modulator-free transmitter}
The encoding is based on the combination of two well-known techniques in laser physics: optical injection locking (OIL)~\cite{LeBinh.2018,Barnes.1993} and direct phase modulation~\cite{Shirasaki.1988}.
This approach has been proven to effectively remove the need for a phase modulator in QKD~\cite{Yuan.2016}: this makes our transmitter versatile while also reducing the number of optical elements in the setup.
Another advantage of our transmitting setup is the lack of an aMZI.
This is useful for two main reasons: first of all, this greatly reduces losses since there is no mismatch between two arms of different length.
In a fibre-based system this does not have a big impact, but chip-based experiments suffer from higher losses, hence a 500~ps or 1~ns delay line would lead to a power mismatch between the two arms.
For this reason, chip-based experiments usually avoid embedding an aMZI in the transmitter, resorting to modulating every pulse with a phase modulator~\cite{Sibson.2017b} or exploiting the directly phase-modulated setup~\cite{Paraiso.2019}.
The second advantage of an aMZI-free setup is the ability to use the same setup to encode different protocols.
This is crucial to our experiment, as it allows us to change different protocols and/or clock rates at will by simply changing the modulation of the driving current, with a transient of less than 5 seconds between every change.

\begin{figure*}[!htbp]
\centering
\includegraphics[width=\linewidth]{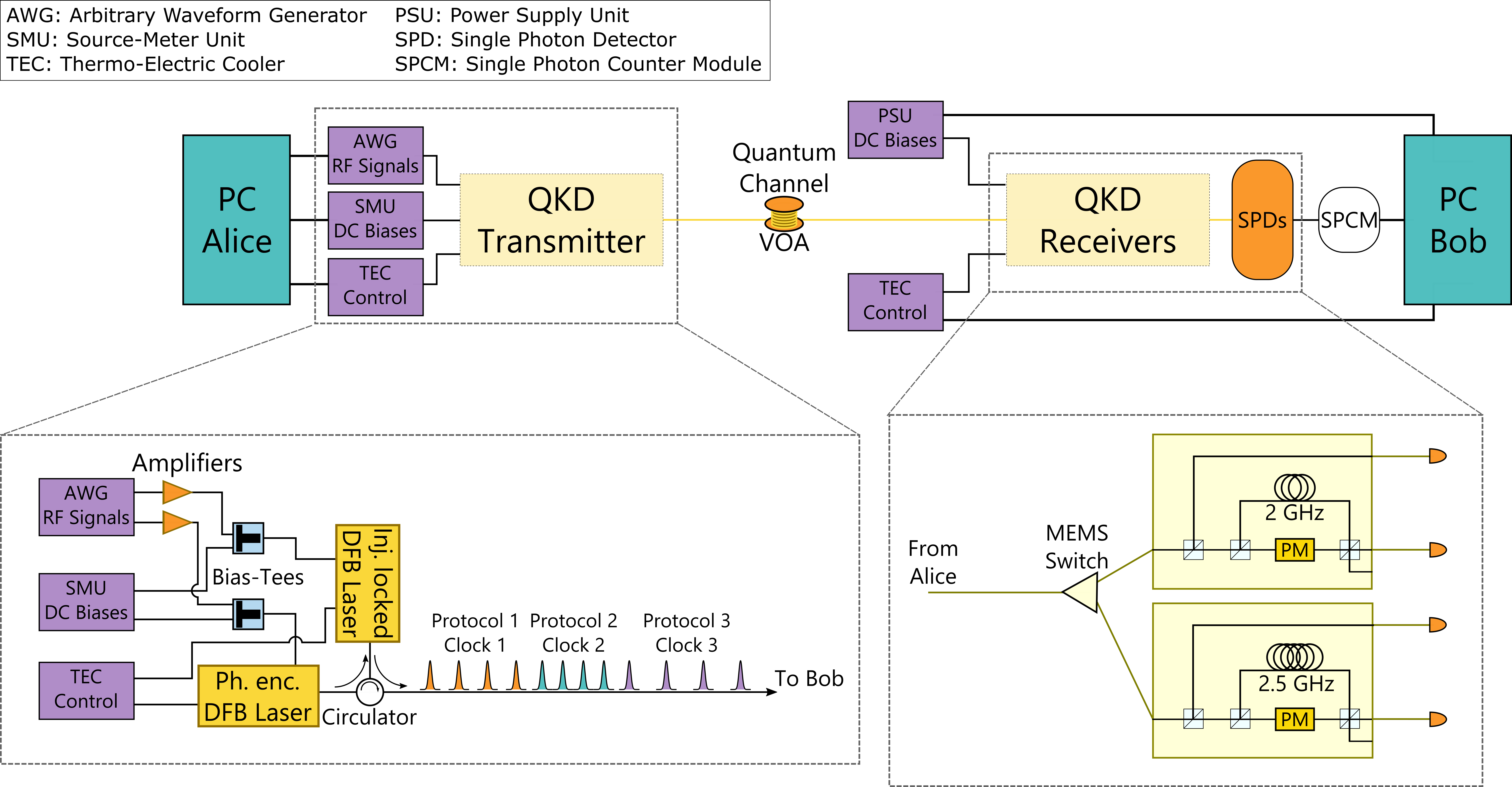}
\caption{\textbf{Experimental setup} Schematic of the experimental setup used to carry out the experiment. DC and RF signals are combined through a bias-tee and sent to the phase encoding and injection locked lasers. A circulator prevents light from the latter going back to the phase encoding laser. The RF signals are changed to encode different protocols and clock rates. An optical MEMS switch selects the correct receiver chip, which is out-coupled to SNSPDs.}
\label{fig:Schematic}
\end{figure*}

The experimental setup is shown in Fig.~\ref{fig:Schematic}.
We use arbitrary waveform generators (AWG) with a sampling rate of 24~Gs/s and vertical resolution of 8 bit to drive the lasers at GHz clock rates and synchronise the system.
A phase encoding laser injects light into a second, injection locked laser through a circulator.
These are both discrete optics, off-the-shelf DFB lasers, with a bandwidth of >10~GHz.
An intensity modulator sets the signal and decoy levels for the system.
The quantum channel is simulated by a variable optical attenuator (VOA), fibre-coupled to the transmitter and the receiver.
A MEMS switch is used to select the correct receiver chip, according to the driving signal sent to the transmitter setup.
At the output of the receiver chips, optical fibres are coupled to superconducting nanowire single photon detectors (SNSPDs)~\cite{Hadfield.2012}; each of the outputs (the time-bin encoding waveguide for COW and one of the interferometer outputs for all three protocols, as mentioned in the previous paragraph) is connected to a detector.
The SNSPDs' output is read by a single-photon counting module (SPCM) which digitises the data and sends the results to Bob's computer where the QBER and key rates are computed.

\begin{figure*}[htbp]
  \centering
  \includegraphics[width=0.9\linewidth]{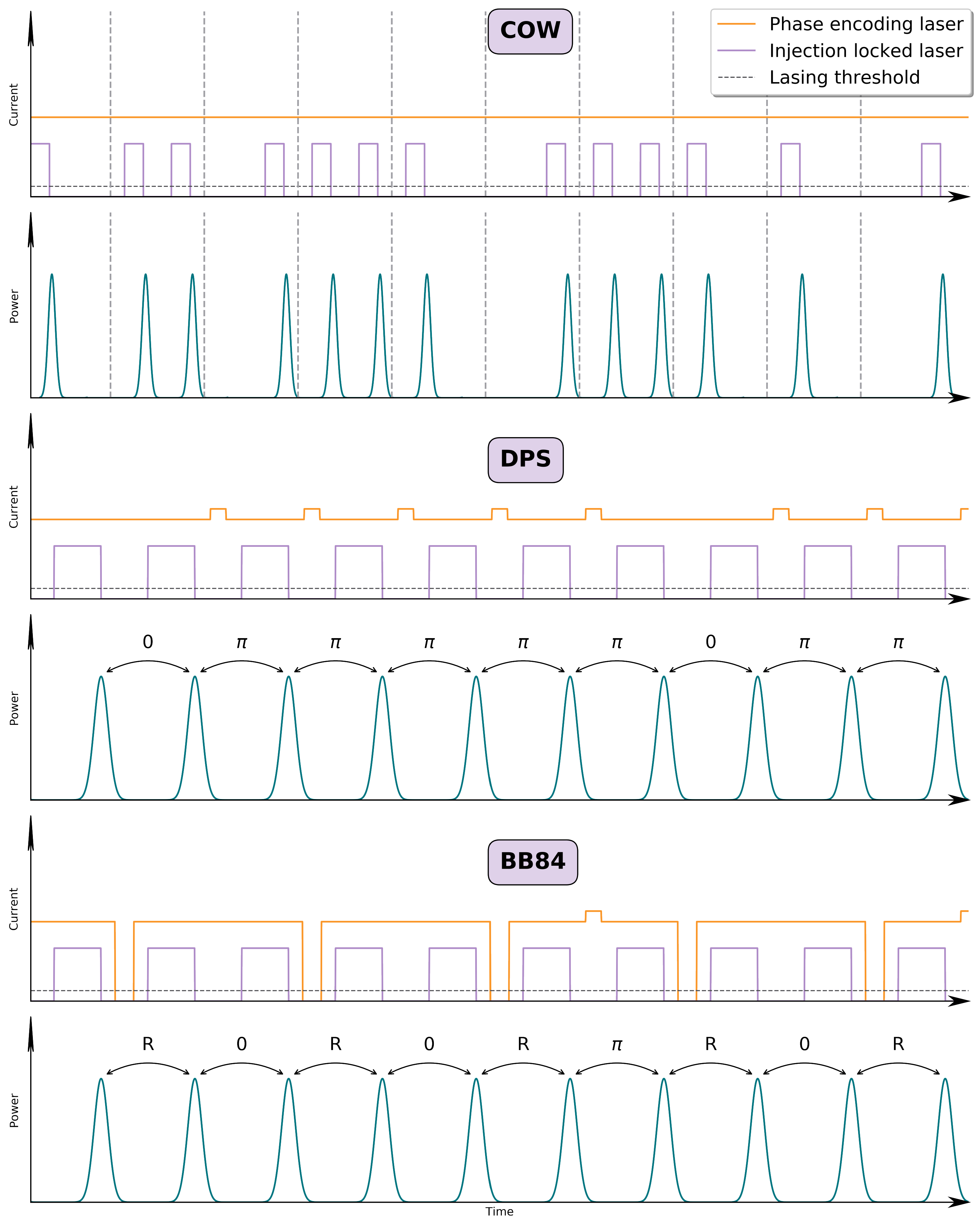}
  \caption{\textbf{Modulation signals} Driving signals from phase encoding and injection locked lasers for the COW, DPS and BB84 protocols.
  For each protocol, the top plot represents the electrical driving signals, the bottom plot represents the optical output. The symbol between consecutive pulses represents the relative phase difference, where ``R'' refers to a random phase difference between pulses.}\label{fig:Modulations}
\end{figure*}

We encode our information as shown in Fig~\ref{fig:Modulations}.
For the COW and DPS protocol, the phase encoding laser is always biased over its threshold to produce continuous-wave (CW) emission;
additionally, in the DPS case, a small current modulation, synchronised to the interval of two injection locked pulses, is applied when a $\pi$ phase modulation is required.
The BB84 protocol requires, in addition to the phase modulation, a random global phase: only pulses belonging to the same pair of time-bins have a set phase difference.
To this end, the phase encoding laser is driven periodically above and below the lasing threshold, exploiting the inherent randomness of gain-switching~\cite{Jofre.2011,Yuan.2014}.

\paragraph*{Integrated receivers}
The receivers are $\mathrm{SiO_xN_y}$ photonic chips, whose interferometer length determine the clock rates they work at.
We use receiver chips whose interferometer delay lengths are 500 ps and 400 ps, achieving clock rates of 2 and 2.5 GHz respectively.

Our QKD receivers are manufactured on a silicon-based substrate, which allows for low losses and easy integration.
Our receiver chips are designed to have different interferometers, each one decoding information for a different protocol.

The interferometers are tuned by means of thermo-optic phase shifters.
Such elements can be driven with a DC current source; a $\pi$ phase shift is obtained with a voltage of around $15~\mathrm{V}$.
The phase shift induced by the heaters can be used to direct light towards one arm or the other of the interferometer, or to balance the power between the two arms.
This is particularly important since the long arm will cause a power imbalance at the output coupler: the length of the delay line will cause more losses.
For this experiment, the total loss for the interferometer circuit on the $2.5~\mathrm{GHz}$ chip is measured to be $10.1~\mathrm{dB}$, while the loss for the $2~\mathrm{GHz}$ chip is $6.7~\mathrm{dB}$.
The straight waveguides used for the COW protocol have losses that are $\sim~3~\mathrm{dB}$ less than the interferometer ones.

The main contribution to the losses is the propagation loss ($\sim 0.2~\mathrm{dB/cm}$) in the delay line, however a big contribution is also coming from fabrication imperfections and fibre coupling.
All these cause the excess loss in the $2.5~\mathrm{GHz}$ chip compared to the $2~\mathrm{GHz}$ chip, as the statistical imperfections of that chip cause losses that overweigh the lower loss coming from the shorter delay line.

The output waveguides are out-coupled through optical fibres to SNSPDs, with an efficiency of $44\%$ and dark count rates of $\sim 10~\mathrm{Hz}$.

\section{Results}

\begin{figure*}[htbp]
\centering
\includegraphics[width=0.9\linewidth]{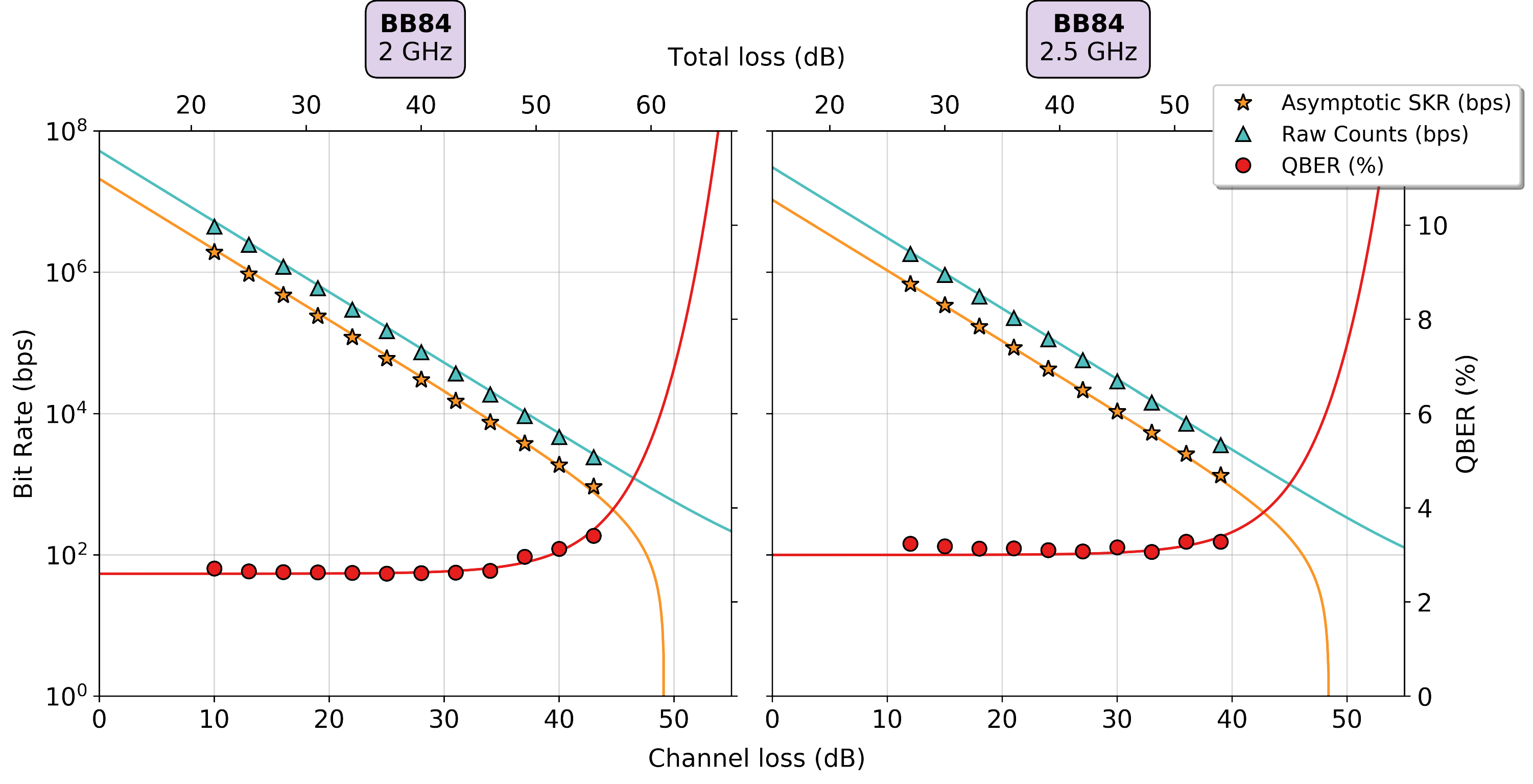}
\caption{\textbf{Secure key rates} BB84 secure key rates and QBER vs channel loss for 2~GHz and 2.5~GHz. The color red indicates the QBER, teal the raw count rates and orange the SKR. The points correspond to measured data, the line to the simulated behaviour.}
\label{fig:SKR}
\end{figure*}

Fig.~\ref{fig:SKR} shows the performance of the transmitter at two different clock rates using the BB84 protocol.
The loss considered in the plots includes only the quantum channel attenuation and the loss from the optical switch.
As shown in the figure, the 2.5~GHz performs worse than the 2~GHz receiver at the same channel loss.
This is due to the excess losses in the chip mentioned in the previous paragraph.
The top axis of Fig.~\ref{fig:SKR} shows instead the total loss, including the receiver chip loss.
This allows a clearer comparison between the chips, and shows how the higher clock rate would indeed yield a higher key rate as expected, if the chips had equal loss.

\begin{figure*}[htbp]
\centering
\includegraphics[width=\linewidth]{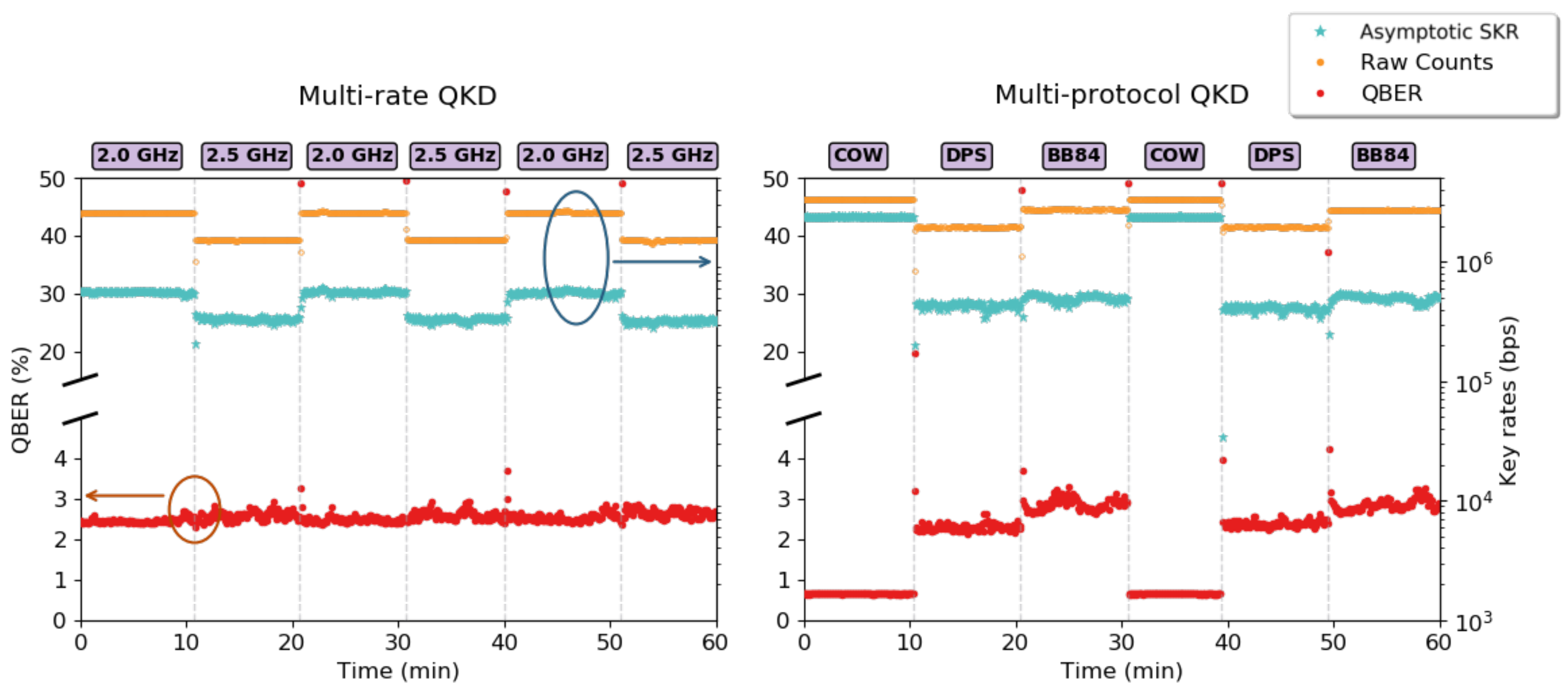}
\caption{\textbf{Performance of the universal transmitter} Results obtained encoding the BB84 protocol at different clock rates (left) and encoding different protocols at a 2 GHz rate (right). The first point after every change of clock rate or protocol, with a high QBER, does not result in a positive SKR. The data is collected at a channel loss of 14~dB. The 2.5~GHz regime has lower count rates than the 2~GHz regime due to the higher losses in the receiver chip.}
\label{fig:switching}
\end{figure*}

We then proceed to the main goal of the paper, demonstrating the flexibility of our system.
Our transmitter is set to implement different protocols at different clock rates.
A signal is then sent to the system, triggering the clock rate and protocol change after 10 minutes.
When this happens, a first point is recorded with a high QBER and, consequently, no positive key rate.
The main reason behind this is that the AWG's electronics takes some time to settle to a stable output.
This time is below 5 seconds.
Results are shown in Fig.~\ref{fig:switching}.
The stability of the system allows for a reasonably constant secure key rate (SKR) during a 10-minute time window, at a channel loss of $14~\mathrm{dB}$.

The QBER is stable around values of $2.7$\% for BB84 at both clock rates and DPS, while it is around $0.7$\% for the COW protocol, due to the absence of phase errors in this protocol.
This yields key rates of $0.5$, $0.4$ and $2.5~\mathrm{Mbps}$ for BB84, DPS and COW, respectively.

\section{Conclusion}

Our experiments show the feasibility of having a single transmitter communicating with receivers employing different clock rates and protocols.
This is possible because of the directly phase-modulated setup which allows flexibility to change the operating conditions by simply modifying the electrical driving signals.
The protocols and clock rates are set up to continuously change, leading to the system reconfiguring in real time while still maintaining a low QBER and high SKR.
We showed an effective setup time lower than $5~\mathrm{s}$.
This was limited by the driving electronics, mainly the AWG, needing time to settle to the desired operation regime.
Reduced times can be certainly obtained by using faster driving electronics, for instance, a custom-designed FPGA board.

On the receiver side, two different chips were used for the two different clock rates.
Tuneable delay length can be considered in order to demonstrate multi-rate capabilities of the receiver, however it must be noted that adding reconfigurable components would add to the complexity of the chip and could cause issues such as thermal fluctuations.
An alternative approach, since the advantage of integrated photonics lies in the compactness of devices, might be to implement different interferometers on the same chip.

Our result is a step forward towards interoperability between devices from different vendors and paves the way for large-scale, collaborative deployment of QKD systems.
In this respect, we believe that our work will have a positive impact on the on-going efforts in QKD standardization.

\section*{Acknowledgments}

I.D.M. acknowledges funding from the European Union’s Horizon 2020 research and innovation programme under the Marie Skłodowska-Curie grant agreement No~675662.

G.L.R. gratefully acknowledges financial support from the EPSRC CDT in Integrated Photonic and Electronics Systems, Toshiba Europe Limited and an industrial fellowship with The Royal Commission for the Exhibition of 1851.

This work has been partially funded by the Innovate UK project AQUASEC, as part of the UK National Quantum Technologies Programme.

\section*{Disclosures}
The authors declare no conflicts of interest.

\section*{Data availability}
 Data underlying the results presented in this paper are not publicly available but may be obtained from the authors upon reasonable request.

\bibliography{RealTimeQTx_IDM_ArXiV}

\begin{thebibliography}{29}%
\makeatletter
\providecommand \@ifxundefined [1]{%
 \@ifx{#1\undefined}
}%
\providecommand \@ifnum [1]{%
 \ifnum #1\expandafter \@firstoftwo
 \else \expandafter \@secondoftwo
 \fi
}%
\providecommand \@ifx [1]{%
 \ifx #1\expandafter \@firstoftwo
 \else \expandafter \@secondoftwo
 \fi
}%
\providecommand \natexlab [1]{#1}%
\providecommand \enquote  [1]{``#1''}%
\providecommand \bibnamefont  [1]{#1}%
\providecommand \bibfnamefont [1]{#1}%
\providecommand \citenamefont [1]{#1}%
\providecommand \href@noop [0]{\@secondoftwo}%
\providecommand \href [0]{\begingroup \@sanitize@url \@href}%
\providecommand \@href[1]{\@@startlink{#1}\@@href}%
\providecommand \@@href[1]{\endgroup#1\@@endlink}%
\providecommand \@sanitize@url [0]{\catcode `\\12\catcode `\$12\catcode
  `\&12\catcode `\#12\catcode `\^12\catcode `\_12\catcode `\%12\relax}%
\providecommand \@@startlink[1]{}%
\providecommand \@@endlink[0]{}%
\providecommand \url  [0]{\begingroup\@sanitize@url \@url }%
\providecommand \@url [1]{\endgroup\@href {#1}{\urlprefix }}%
\providecommand \urlprefix  [0]{URL }%
\providecommand \Eprint [0]{\href }%
\providecommand \doibase [0]{https://doi.org/}%
\providecommand \selectlanguage [0]{\@gobble}%
\providecommand \bibinfo  [0]{\@secondoftwo}%
\providecommand \bibfield  [0]{\@secondoftwo}%
\providecommand \translation [1]{[#1]}%
\providecommand \BibitemOpen [0]{}%
\providecommand \bibitemStop [0]{}%
\providecommand \bibitemNoStop [0]{.\EOS\space}%
\providecommand \EOS [0]{\spacefactor3000\relax}%
\providecommand \BibitemShut  [1]{\csname bibitem#1\endcsname}%
\let\auto@bib@innerbib\@empty
\bibitem [{\citenamefont {Scarani}\ \emph {et~al.}(2009)\citenamefont
  {Scarani}, \citenamefont {Bechmann-Pasquinucci}, \citenamefont {Cerf},
  \citenamefont {Du{\v{s}}ek}, \citenamefont {L{\"u}tkenhaus},\ and\
  \citenamefont {Peev}}]{Scarani.2009}%
  \BibitemOpen
  \bibfield  {author} {\bibinfo {author} {\bibfnamefont {V.}~\bibnamefont
  {Scarani}}, \bibinfo {author} {\bibfnamefont {H.}~\bibnamefont
  {Bechmann-Pasquinucci}}, \bibinfo {author} {\bibfnamefont {N.~J.}\
  \bibnamefont {Cerf}}, \bibinfo {author} {\bibfnamefont {M.}~\bibnamefont
  {Du{\v{s}}ek}}, \bibinfo {author} {\bibfnamefont {N.}~\bibnamefont
  {L{\"u}tkenhaus}},\ and\ \bibinfo {author} {\bibfnamefont {M.}~\bibnamefont
  {Peev}},\ }\bibfield  {title} {\bibinfo {title} {The security of practical
  quantum key distribution},\ }\href@noop {} {\bibfield  {journal} {\bibinfo
  {journal} {Rev. Mod. Phys.}\ }\textbf {\bibinfo {volume} {81}},\ \bibinfo
  {pages} {1301} (\bibinfo {year} {2009})}\BibitemShut {NoStop}%
\bibitem [{\citenamefont {Arute}\ \emph {et~al.}(2019)\citenamefont {Arute},
  \citenamefont {Arya}, \citenamefont {Babbush}, \citenamefont {Bacon},
  \citenamefont {Bardin}, \citenamefont {Barends}, \citenamefont {Biswas},
  \citenamefont {Boixo}, \citenamefont {Brandao}, \citenamefont {Buell},
  \citenamefont {Burkett}, \citenamefont {Chen}, \citenamefont {Chen},
  \citenamefont {Chiaro}, \citenamefont {Collins}, \citenamefont {Courtney},
  \citenamefont {Dunsworth}, \citenamefont {Farhi}, \citenamefont {Foxen},
  \citenamefont {Fowler}, \citenamefont {Gidney}, \citenamefont {Giustina},
  \citenamefont {Graff}, \citenamefont {Guerin}, \citenamefont {Habegger},
  \citenamefont {Harrigan}, \citenamefont {Hartmann}, \citenamefont {Ho},
  \citenamefont {Hoffmann}, \citenamefont {Huang}, \citenamefont {Humble},
  \citenamefont {Isakov}, \citenamefont {Jeffrey}, \citenamefont {Jiang},
  \citenamefont {Kafri}, \citenamefont {Kechedzhi}, \citenamefont {Kelly},
  \citenamefont {Klimov}, \citenamefont {Knysh}, \citenamefont {Korotkov},
  \citenamefont {Kostritsa}, \citenamefont {Landhuis}, \citenamefont
  {Lindmark}, \citenamefont {Lucero}, \citenamefont {Lyakh}, \citenamefont
  {Mandr{\`a}}, \citenamefont {McClean}, \citenamefont {McEwen}, \citenamefont
  {Megrant}, \citenamefont {Mi}, \citenamefont {Michielsen}, \citenamefont
  {Mohseni}, \citenamefont {Mutus}, \citenamefont {Naaman}, \citenamefont
  {Neeley}, \citenamefont {Neill}, \citenamefont {Niu}, \citenamefont {Ostby},
  \citenamefont {Petukhov}, \citenamefont {Platt}, \citenamefont {Quintana},
  \citenamefont {Rieffel}, \citenamefont {Roushan}, \citenamefont {Rubin},
  \citenamefont {Sank}, \citenamefont {Satzinger}, \citenamefont {Smelyanskiy},
  \citenamefont {Sung}, \citenamefont {Trevithick}, \citenamefont
  {Vainsencher}, \citenamefont {Villalonga}, \citenamefont {White},
  \citenamefont {Yao}, \citenamefont {Yeh}, \citenamefont {Zalcman},
  \citenamefont {Neven},\ and\ \citenamefont {Martinis}}]{Arute.2019}%
  \BibitemOpen
  \bibfield  {author} {\bibinfo {author} {\bibfnamefont {F.}~\bibnamefont
  {Arute}}, \bibinfo {author} {\bibfnamefont {K.}~\bibnamefont {Arya}},
  \bibinfo {author} {\bibfnamefont {R.}~\bibnamefont {Babbush}}, \bibinfo
  {author} {\bibfnamefont {D.}~\bibnamefont {Bacon}}, \bibinfo {author}
  {\bibfnamefont {J.~C.}\ \bibnamefont {Bardin}}, \bibinfo {author}
  {\bibfnamefont {R.}~\bibnamefont {Barends}}, \bibinfo {author} {\bibfnamefont
  {R.}~\bibnamefont {Biswas}}, \bibinfo {author} {\bibfnamefont
  {S.}~\bibnamefont {Boixo}}, \bibinfo {author} {\bibfnamefont {F.~G. S.~L.}\
  \bibnamefont {Brandao}}, \bibinfo {author} {\bibfnamefont {D.~A.}\
  \bibnamefont {Buell}}, \bibinfo {author} {\bibfnamefont {B.}~\bibnamefont
  {Burkett}}, \bibinfo {author} {\bibfnamefont {Y.}~\bibnamefont {Chen}},
  \bibinfo {author} {\bibfnamefont {Z.}~\bibnamefont {Chen}}, \bibinfo {author}
  {\bibfnamefont {B.}~\bibnamefont {Chiaro}}, \bibinfo {author} {\bibfnamefont
  {R.}~\bibnamefont {Collins}}, \bibinfo {author} {\bibfnamefont
  {W.}~\bibnamefont {Courtney}}, \bibinfo {author} {\bibfnamefont
  {A.}~\bibnamefont {Dunsworth}}, \bibinfo {author} {\bibfnamefont
  {E.}~\bibnamefont {Farhi}}, \bibinfo {author} {\bibfnamefont
  {B.}~\bibnamefont {Foxen}}, \bibinfo {author} {\bibfnamefont
  {A.}~\bibnamefont {Fowler}}, \bibinfo {author} {\bibfnamefont
  {C.}~\bibnamefont {Gidney}}, \bibinfo {author} {\bibfnamefont
  {M.}~\bibnamefont {Giustina}}, \bibinfo {author} {\bibfnamefont
  {R.}~\bibnamefont {Graff}}, \bibinfo {author} {\bibfnamefont
  {K.}~\bibnamefont {Guerin}}, \bibinfo {author} {\bibfnamefont
  {S.}~\bibnamefont {Habegger}}, \bibinfo {author} {\bibfnamefont {M.~P.}\
  \bibnamefont {Harrigan}}, \bibinfo {author} {\bibfnamefont {M.~J.}\
  \bibnamefont {Hartmann}}, \bibinfo {author} {\bibfnamefont {A.}~\bibnamefont
  {Ho}}, \bibinfo {author} {\bibfnamefont {M.}~\bibnamefont {Hoffmann}},
  \bibinfo {author} {\bibfnamefont {T.}~\bibnamefont {Huang}}, \bibinfo
  {author} {\bibfnamefont {T.~S.}\ \bibnamefont {Humble}}, \bibinfo {author}
  {\bibfnamefont {S.~V.}\ \bibnamefont {Isakov}}, \bibinfo {author}
  {\bibfnamefont {E.}~\bibnamefont {Jeffrey}}, \bibinfo {author} {\bibfnamefont
  {Z.}~\bibnamefont {Jiang}}, \bibinfo {author} {\bibfnamefont
  {D.}~\bibnamefont {Kafri}}, \bibinfo {author} {\bibfnamefont
  {K.}~\bibnamefont {Kechedzhi}}, \bibinfo {author} {\bibfnamefont
  {J.}~\bibnamefont {Kelly}}, \bibinfo {author} {\bibfnamefont {P.~V.}\
  \bibnamefont {Klimov}}, \bibinfo {author} {\bibfnamefont {S.}~\bibnamefont
  {Knysh}}, \bibinfo {author} {\bibfnamefont {A.}~\bibnamefont {Korotkov}},
  \bibinfo {author} {\bibfnamefont {F.}~\bibnamefont {Kostritsa}}, \bibinfo
  {author} {\bibfnamefont {D.}~\bibnamefont {Landhuis}}, \bibinfo {author}
  {\bibfnamefont {M.}~\bibnamefont {Lindmark}}, \bibinfo {author}
  {\bibfnamefont {E.}~\bibnamefont {Lucero}}, \bibinfo {author} {\bibfnamefont
  {D.}~\bibnamefont {Lyakh}}, \bibinfo {author} {\bibfnamefont
  {S.}~\bibnamefont {Mandr{\`a}}}, \bibinfo {author} {\bibfnamefont {J.~R.}\
  \bibnamefont {McClean}}, \bibinfo {author} {\bibfnamefont {M.}~\bibnamefont
  {McEwen}}, \bibinfo {author} {\bibfnamefont {A.}~\bibnamefont {Megrant}},
  \bibinfo {author} {\bibfnamefont {X.}~\bibnamefont {Mi}}, \bibinfo {author}
  {\bibfnamefont {K.}~\bibnamefont {Michielsen}}, \bibinfo {author}
  {\bibfnamefont {M.}~\bibnamefont {Mohseni}}, \bibinfo {author} {\bibfnamefont
  {J.}~\bibnamefont {Mutus}}, \bibinfo {author} {\bibfnamefont
  {O.}~\bibnamefont {Naaman}}, \bibinfo {author} {\bibfnamefont
  {M.}~\bibnamefont {Neeley}}, \bibinfo {author} {\bibfnamefont
  {C.}~\bibnamefont {Neill}}, \bibinfo {author} {\bibfnamefont {M.~Y.}\
  \bibnamefont {Niu}}, \bibinfo {author} {\bibfnamefont {E.}~\bibnamefont
  {Ostby}}, \bibinfo {author} {\bibfnamefont {A.}~\bibnamefont {Petukhov}},
  \bibinfo {author} {\bibfnamefont {J.~C.}\ \bibnamefont {Platt}}, \bibinfo
  {author} {\bibfnamefont {C.}~\bibnamefont {Quintana}}, \bibinfo {author}
  {\bibfnamefont {E.~G.}\ \bibnamefont {Rieffel}}, \bibinfo {author}
  {\bibfnamefont {P.}~\bibnamefont {Roushan}}, \bibinfo {author} {\bibfnamefont
  {N.~C.}\ \bibnamefont {Rubin}}, \bibinfo {author} {\bibfnamefont
  {D.}~\bibnamefont {Sank}}, \bibinfo {author} {\bibfnamefont {K.~J.}\
  \bibnamefont {Satzinger}}, \bibinfo {author} {\bibfnamefont {V.}~\bibnamefont
  {Smelyanskiy}}, \bibinfo {author} {\bibfnamefont {K.~J.}\ \bibnamefont
  {Sung}}, \bibinfo {author} {\bibfnamefont {M.~D.}\ \bibnamefont
  {Trevithick}}, \bibinfo {author} {\bibfnamefont {A.}~\bibnamefont
  {Vainsencher}}, \bibinfo {author} {\bibfnamefont {B.}~\bibnamefont
  {Villalonga}}, \bibinfo {author} {\bibfnamefont {T.}~\bibnamefont {White}},
  \bibinfo {author} {\bibfnamefont {Z.~J.}\ \bibnamefont {Yao}}, \bibinfo
  {author} {\bibfnamefont {P.}~\bibnamefont {Yeh}}, \bibinfo {author}
  {\bibfnamefont {A.}~\bibnamefont {Zalcman}}, \bibinfo {author} {\bibfnamefont
  {H.}~\bibnamefont {Neven}},\ and\ \bibinfo {author} {\bibfnamefont {J.~M.}\
  \bibnamefont {Martinis}},\ }\bibfield  {title} {\bibinfo {title} {Quantum
  supremacy using a programmable superconducting processor},\ }\href
  {https://doi.org/10.1038/s41586-019-1666-5} {\bibfield  {journal} {\bibinfo
  {journal} {Nature}\ }\textbf {\bibinfo {volume} {574}},\ \bibinfo {pages}
  {505} (\bibinfo {year} {2019})}\BibitemShut {NoStop}%
\bibitem [{\citenamefont {Shor}(1999)}]{Shor.1999}%
  \BibitemOpen
  \bibfield  {author} {\bibinfo {author} {\bibfnamefont {P.~W.}\ \bibnamefont
  {Shor}},\ }\bibfield  {title} {\bibinfo {title} {Polynomial-time algorithms
  for prime factorization and discrete logarithms on a quantum computer},\
  }\href@noop {} {\bibfield  {journal} {\bibinfo  {journal} {SIAM review}\
  }\textbf {\bibinfo {volume} {41}},\ \bibinfo {pages} {303} (\bibinfo {year}
  {1999})}\BibitemShut {NoStop}%
\bibitem [{\citenamefont {Yuan}\ \emph {et~al.}(2018)\citenamefont {Yuan},
  \citenamefont {Plews}, \citenamefont {Takahashi}, \citenamefont {Doi},
  \citenamefont {Tam}, \citenamefont {Sharpe}, \citenamefont {Dixon},
  \citenamefont {Lavelle}, \citenamefont {Dynes}, \citenamefont {Murakami},
  \citenamefont {Kujiraoka}, \citenamefont {Lucamarini}, \citenamefont
  {Tanizawa}, \citenamefont {Sato},\ and\ \citenamefont {Shields}}]{Yuan.2018}%
  \BibitemOpen
  \bibfield  {author} {\bibinfo {author} {\bibfnamefont {Z.}~\bibnamefont
  {Yuan}}, \bibinfo {author} {\bibfnamefont {A.}~\bibnamefont {Plews}},
  \bibinfo {author} {\bibfnamefont {R.}~\bibnamefont {Takahashi}}, \bibinfo
  {author} {\bibfnamefont {K.}~\bibnamefont {Doi}}, \bibinfo {author}
  {\bibfnamefont {W.}~\bibnamefont {Tam}}, \bibinfo {author} {\bibfnamefont
  {A.}~\bibnamefont {Sharpe}}, \bibinfo {author} {\bibfnamefont
  {A.}~\bibnamefont {Dixon}}, \bibinfo {author} {\bibfnamefont
  {E.}~\bibnamefont {Lavelle}}, \bibinfo {author} {\bibfnamefont
  {J.}~\bibnamefont {Dynes}}, \bibinfo {author} {\bibfnamefont
  {A.}~\bibnamefont {Murakami}}, \bibinfo {author} {\bibfnamefont
  {M.}~\bibnamefont {Kujiraoka}}, \bibinfo {author} {\bibfnamefont
  {M.}~\bibnamefont {Lucamarini}}, \bibinfo {author} {\bibfnamefont
  {Y.}~\bibnamefont {Tanizawa}}, \bibinfo {author} {\bibfnamefont
  {H.}~\bibnamefont {Sato}},\ and\ \bibinfo {author} {\bibfnamefont {A.~J.}\
  \bibnamefont {Shields}},\ }\bibfield  {title} {\bibinfo {title} {10-mb/s
  quantum key distribution},\ }\href@noop {} {\bibfield  {journal} {\bibinfo
  {journal} {Journal of Lightwave Technology}\ }\textbf {\bibinfo {volume}
  {36}},\ \bibinfo {pages} {3427} (\bibinfo {year} {2018})}\BibitemShut
  {NoStop}%
\bibitem [{\citenamefont {Boaron}\ \emph {et~al.}(2018)\citenamefont {Boaron},
  \citenamefont {Korzh}, \citenamefont {Houlmann}, \citenamefont {Boso},
  \citenamefont {Rusca}, \citenamefont {Gray}, \citenamefont {Li},
  \citenamefont {Nolan}, \citenamefont {Martin},\ and\ \citenamefont
  {Zbinden}}]{Boaron.2018}%
  \BibitemOpen
  \bibfield  {author} {\bibinfo {author} {\bibfnamefont {A.}~\bibnamefont
  {Boaron}}, \bibinfo {author} {\bibfnamefont {B.}~\bibnamefont {Korzh}},
  \bibinfo {author} {\bibfnamefont {R.}~\bibnamefont {Houlmann}}, \bibinfo
  {author} {\bibfnamefont {G.}~\bibnamefont {Boso}}, \bibinfo {author}
  {\bibfnamefont {D.}~\bibnamefont {Rusca}}, \bibinfo {author} {\bibfnamefont
  {S.}~\bibnamefont {Gray}}, \bibinfo {author} {\bibfnamefont {M.-J.}\
  \bibnamefont {Li}}, \bibinfo {author} {\bibfnamefont {D.}~\bibnamefont
  {Nolan}}, \bibinfo {author} {\bibfnamefont {A.}~\bibnamefont {Martin}},\ and\
  \bibinfo {author} {\bibfnamefont {H.}~\bibnamefont {Zbinden}},\ }\bibfield
  {title} {\bibinfo {title} {Simple 2.5 ghz time-bin quantum key
  distribution},\ }\href {https://doi.org/10.1063/1.5027030} {\bibfield
  {journal} {\bibinfo  {journal} {Applied Physics Letters}\ }\textbf {\bibinfo
  {volume} {112}},\ \bibinfo {pages} {171108} (\bibinfo {year}
  {2018})}\BibitemShut {NoStop}%
\bibitem [{\citenamefont {Lucamarini}\ \emph {et~al.}(2018)\citenamefont
  {Lucamarini}, \citenamefont {Yuan}, \citenamefont {Dynes},\ and\
  \citenamefont {Shields}}]{Lucamarini.2018}%
  \BibitemOpen
  \bibfield  {author} {\bibinfo {author} {\bibfnamefont {M.}~\bibnamefont
  {Lucamarini}}, \bibinfo {author} {\bibfnamefont {Z.~L.}\ \bibnamefont
  {Yuan}}, \bibinfo {author} {\bibfnamefont {J.~F.}\ \bibnamefont {Dynes}},\
  and\ \bibinfo {author} {\bibfnamefont {A.~J.}\ \bibnamefont {Shields}},\
  }\bibfield  {title} {\bibinfo {title} {Overcoming the rate--distance limit of
  quantum key distribution without quantum repeaters},\ }\href
  {https://doi.org/10.1038/s41586-018-0066-6} {\bibfield  {journal} {\bibinfo
  {journal} {Nature}\ }\textbf {\bibinfo {volume} {557}},\ \bibinfo {pages}
  {400} (\bibinfo {year} {2018})}\BibitemShut {NoStop}%
\bibitem [{\citenamefont {Minder}\ \emph {et~al.}(2019)\citenamefont {Minder},
  \citenamefont {Pittaluga}, \citenamefont {Roberts}, \citenamefont
  {Lucamarini}, \citenamefont {Dynes}, \citenamefont {Yuan},\ and\
  \citenamefont {Shields}}]{Minder.2019}%
  \BibitemOpen
  \bibfield  {author} {\bibinfo {author} {\bibfnamefont {M.}~\bibnamefont
  {Minder}}, \bibinfo {author} {\bibfnamefont {M.}~\bibnamefont {Pittaluga}},
  \bibinfo {author} {\bibfnamefont {G.~L.}\ \bibnamefont {Roberts}}, \bibinfo
  {author} {\bibfnamefont {M.}~\bibnamefont {Lucamarini}}, \bibinfo {author}
  {\bibfnamefont {J.~F.}\ \bibnamefont {Dynes}}, \bibinfo {author}
  {\bibfnamefont {Z.~L.}\ \bibnamefont {Yuan}},\ and\ \bibinfo {author}
  {\bibfnamefont {A.~J.}\ \bibnamefont {Shields}},\ }\bibfield  {title}
  {\bibinfo {title} {Experimental quantum key distribution beyond the
  repeaterless secret key capacity},\ }\href
  {https://doi.org/10.1038/s41566-019-0377-7} {\bibfield  {journal} {\bibinfo
  {journal} {Nature Photonics}\ }\textbf {\bibinfo {volume} {13}},\ \bibinfo
  {pages} {334} (\bibinfo {year} {2019})}\BibitemShut {NoStop}%
\bibitem [{\citenamefont {Fang}\ \emph {et~al.}(2020)\citenamefont {Fang},
  \citenamefont {Zeng}, \citenamefont {Liu}, \citenamefont {Zou}, \citenamefont
  {Wu}, \citenamefont {Tang}, \citenamefont {Sheng}, \citenamefont {Xiang},
  \citenamefont {Zhang}, \citenamefont {Li}, \citenamefont {Wang},
  \citenamefont {You}, \citenamefont {Li}, \citenamefont {Chen}, \citenamefont
  {Chen}, \citenamefont {Zhang}, \citenamefont {Peng}, \citenamefont {Ma},
  \citenamefont {Chen},\ and\ \citenamefont {Pan}}]{Fang.2020}%
  \BibitemOpen
  \bibfield  {author} {\bibinfo {author} {\bibfnamefont {X.-T.}\ \bibnamefont
  {Fang}}, \bibinfo {author} {\bibfnamefont {P.}~\bibnamefont {Zeng}}, \bibinfo
  {author} {\bibfnamefont {H.}~\bibnamefont {Liu}}, \bibinfo {author}
  {\bibfnamefont {M.}~\bibnamefont {Zou}}, \bibinfo {author} {\bibfnamefont
  {W.}~\bibnamefont {Wu}}, \bibinfo {author} {\bibfnamefont {Y.-L.}\
  \bibnamefont {Tang}}, \bibinfo {author} {\bibfnamefont {Y.-J.}\ \bibnamefont
  {Sheng}}, \bibinfo {author} {\bibfnamefont {Y.}~\bibnamefont {Xiang}},
  \bibinfo {author} {\bibfnamefont {W.}~\bibnamefont {Zhang}}, \bibinfo
  {author} {\bibfnamefont {H.}~\bibnamefont {Li}}, \bibinfo {author}
  {\bibfnamefont {Z.}~\bibnamefont {Wang}}, \bibinfo {author} {\bibfnamefont
  {L.}~\bibnamefont {You}}, \bibinfo {author} {\bibfnamefont {M.-J.}\
  \bibnamefont {Li}}, \bibinfo {author} {\bibfnamefont {H.}~\bibnamefont
  {Chen}}, \bibinfo {author} {\bibfnamefont {Y.-A.}\ \bibnamefont {Chen}},
  \bibinfo {author} {\bibfnamefont {Q.}~\bibnamefont {Zhang}}, \bibinfo
  {author} {\bibfnamefont {C.-Z.}\ \bibnamefont {Peng}}, \bibinfo {author}
  {\bibfnamefont {X.}~\bibnamefont {Ma}}, \bibinfo {author} {\bibfnamefont
  {T.-Y.}\ \bibnamefont {Chen}},\ and\ \bibinfo {author} {\bibfnamefont
  {J.-W.}\ \bibnamefont {Pan}},\ }\bibfield  {title} {\bibinfo {title}
  {Implementation of quantum key distribution surpassing the linear
  rate-transmittance bound},\ }\href
  {https://doi.org/10.1038/s41566-020-0599-8} {\bibfield  {journal} {\bibinfo
  {journal} {Nature Photonics}\ }\textbf {\bibinfo {volume} {14}},\ \bibinfo
  {pages} {422} (\bibinfo {year} {2020})}\BibitemShut {NoStop}%
\bibitem [{\citenamefont {Moseley}\ \emph {et~al.}(2004)\citenamefont
  {Moseley}, \citenamefont {Randall},\ and\ \citenamefont
  {Wiles}}]{Moseley.2004}%
  \BibitemOpen
  \bibfield  {author} {\bibinfo {author} {\bibfnamefont {S.}~\bibnamefont
  {Moseley}}, \bibinfo {author} {\bibfnamefont {S.}~\bibnamefont {Randall}},\
  and\ \bibinfo {author} {\bibfnamefont {A.}~\bibnamefont {Wiles}},\ }\bibfield
   {title} {\bibinfo {title} {In pursuit of interoperability},\ }\href
  {https://doi.org/10.4018/jitsr.2004070103} {\bibfield  {journal} {\bibinfo
  {journal} {International Journal of IT Standards and Standardization
  Research}\ }\textbf {\bibinfo {volume} {2}},\ \bibinfo {pages} {34} (\bibinfo
  {year} {2004})}\BibitemShut {NoStop}%
\bibitem [{\citenamefont {Korzh}\ \emph {et~al.}(2013)\citenamefont {Korzh},
  \citenamefont {Walenta}, \citenamefont {Houlmann},\ and\ \citenamefont
  {Zbinden}}]{Korzh.2013}%
  \BibitemOpen
  \bibfield  {author} {\bibinfo {author} {\bibfnamefont {B.}~\bibnamefont
  {Korzh}}, \bibinfo {author} {\bibfnamefont {N.}~\bibnamefont {Walenta}},
  \bibinfo {author} {\bibfnamefont {R.}~\bibnamefont {Houlmann}},\ and\
  \bibinfo {author} {\bibfnamefont {H.}~\bibnamefont {Zbinden}},\ }\bibfield
  {title} {\bibinfo {title} {A high-speed multi-protocol quantum key
  distribution transmitter based on a dual-drive modulator},\ }\href
  {https://doi.org/10.1364/OE.21.019579} {\bibfield  {journal} {\bibinfo
  {journal} {Optics Express}\ }\textbf {\bibinfo {volume} {21}},\ \bibinfo
  {pages} {19579} (\bibinfo {year} {2013})}\BibitemShut {NoStop}%
\bibitem [{\citenamefont {Sibson}\ \emph {et~al.}(2017)\citenamefont {Sibson},
  \citenamefont {Kennard}, \citenamefont {Stanisic}, \citenamefont {Erven},
  \citenamefont {O'Brien},\ and\ \citenamefont {Thompson}}]{Sibson.2017b}%
  \BibitemOpen
  \bibfield  {author} {\bibinfo {author} {\bibfnamefont {P.}~\bibnamefont
  {Sibson}}, \bibinfo {author} {\bibfnamefont {J.~E.}\ \bibnamefont {Kennard}},
  \bibinfo {author} {\bibfnamefont {S.}~\bibnamefont {Stanisic}}, \bibinfo
  {author} {\bibfnamefont {C.}~\bibnamefont {Erven}}, \bibinfo {author}
  {\bibfnamefont {J.~L.}\ \bibnamefont {O'Brien}},\ and\ \bibinfo {author}
  {\bibfnamefont {M.~G.}\ \bibnamefont {Thompson}},\ }\bibfield  {title}
  {\bibinfo {title} {Integrated silicon photonics for high-speed quantum key
  distribution},\ }\href@noop {} {\bibfield  {journal} {\bibinfo  {journal}
  {Optica}\ }\textbf {\bibinfo {volume} {4}},\ \bibinfo {pages} {172} (\bibinfo
  {year} {2017})}\BibitemShut {NoStop}%
\bibitem [{\citenamefont {Yuan}\ \emph {et~al.}(2016)\citenamefont {Yuan},
  \citenamefont {Fr{\"o}hlich}, \citenamefont {Lucamarini}, \citenamefont
  {Roberts}, \citenamefont {Dynes},\ and\ \citenamefont {Shields}}]{Yuan.2016}%
  \BibitemOpen
  \bibfield  {author} {\bibinfo {author} {\bibfnamefont {Z.~L.}\ \bibnamefont
  {Yuan}}, \bibinfo {author} {\bibfnamefont {B.}~\bibnamefont {Fr{\"o}hlich}},
  \bibinfo {author} {\bibfnamefont {M.}~\bibnamefont {Lucamarini}}, \bibinfo
  {author} {\bibfnamefont {G.~L.}\ \bibnamefont {Roberts}}, \bibinfo {author}
  {\bibfnamefont {J.~F.}\ \bibnamefont {Dynes}},\ and\ \bibinfo {author}
  {\bibfnamefont {A.~J.}\ \bibnamefont {Shields}},\ }\bibfield  {title}
  {\bibinfo {title} {Directly phase-modulated light source},\ }\href
  {https://doi.org/10.1103/PhysRevX.6.031044} {\bibfield  {journal} {\bibinfo
  {journal} {Physical Review X}\ }\textbf {\bibinfo {volume} {6}},\ \bibinfo
  {pages} {031044} (\bibinfo {year} {2016})}\BibitemShut {NoStop}%
\bibitem [{\citenamefont {Para{\"i}so}\ \emph {et~al.}(2019)\citenamefont
  {Para{\"i}so}, \citenamefont {{De Marco}}, \citenamefont {Roger},
  \citenamefont {Marangon}, \citenamefont {Dynes}, \citenamefont {Lucamarini},
  \citenamefont {Yuan},\ and\ \citenamefont {Shields}}]{Paraiso.2019}%
  \BibitemOpen
  \bibfield  {author} {\bibinfo {author} {\bibfnamefont {T.~K.}\ \bibnamefont
  {Para{\"i}so}}, \bibinfo {author} {\bibfnamefont {I.}~\bibnamefont {{De
  Marco}}}, \bibinfo {author} {\bibfnamefont {T.}~\bibnamefont {Roger}},
  \bibinfo {author} {\bibfnamefont {D.~G.}\ \bibnamefont {Marangon}}, \bibinfo
  {author} {\bibfnamefont {J.~F.}\ \bibnamefont {Dynes}}, \bibinfo {author}
  {\bibfnamefont {M.}~\bibnamefont {Lucamarini}}, \bibinfo {author}
  {\bibfnamefont {Z.}~\bibnamefont {Yuan}},\ and\ \bibinfo {author}
  {\bibfnamefont {A.~J.}\ \bibnamefont {Shields}},\ }\bibfield  {title}
  {\bibinfo {title} {A modulator-free quantum key distribution transmitter
  chip},\ }\href {https://doi.org/10.1038/s41534-019-0158-7} {\bibfield
  {journal} {\bibinfo  {journal} {npj Quantum Information}\ }\textbf {\bibinfo
  {volume} {5}},\ \bibinfo {pages} {42} (\bibinfo {year} {2019})}\BibitemShut
  {NoStop}%
\bibitem [{\citenamefont {Inoue}\ \emph {et~al.}(2003)\citenamefont {Inoue},
  \citenamefont {Waks},\ and\ \citenamefont {Yamamoto}}]{Inoue.2003}%
  \BibitemOpen
  \bibfield  {author} {\bibinfo {author} {\bibfnamefont {K.}~\bibnamefont
  {Inoue}}, \bibinfo {author} {\bibfnamefont {E.}~\bibnamefont {Waks}},\ and\
  \bibinfo {author} {\bibfnamefont {Y.}~\bibnamefont {Yamamoto}},\ }\bibfield
  {title} {\bibinfo {title} {Differential-phase-shift quantum key distribution
  using coherent light},\ }\href {https://doi.org/10.1103/PhysRevA.68.022317}
  {\bibfield  {journal} {\bibinfo  {journal} {Physical Review A}\ }\textbf
  {\bibinfo {volume} {68}},\ \bibinfo {pages} {022317} (\bibinfo {year}
  {2003})}\BibitemShut {NoStop}%
\bibitem [{\citenamefont {Bennett}\ and\ \citenamefont
  {Brassard}(1984)}]{Bennett.1984}%
  \BibitemOpen
  \bibfield  {author} {\bibinfo {author} {\bibfnamefont {C.~H.}\ \bibnamefont
  {Bennett}}\ and\ \bibinfo {author} {\bibfnamefont {G.}~\bibnamefont
  {Brassard}},\ }\bibfield  {title} {\bibinfo {title} {Quantum cryptography:
  Public key distribution and coin tossing},\ }in\ \href@noop {} {\emph
  {\bibinfo {booktitle} {International Conference on Computer System and Signal
  Processing, IEEE, 1984}}}\ (\bibinfo {year} {1984})\ pp.\ \bibinfo {pages}
  {175--179}\BibitemShut {NoStop}%
\bibitem [{\citenamefont {Lo}\ \emph {et~al.}(2005)\citenamefont {Lo},
  \citenamefont {Ma},\ and\ \citenamefont {Chen}}]{Lo.2005}%
  \BibitemOpen
  \bibfield  {author} {\bibinfo {author} {\bibfnamefont {H.-K.}\ \bibnamefont
  {Lo}}, \bibinfo {author} {\bibfnamefont {X.}~\bibnamefont {Ma}},\ and\
  \bibinfo {author} {\bibfnamefont {K.}~\bibnamefont {Chen}},\ }\bibfield
  {title} {\bibinfo {title} {Decoy state quantum key distribution},\ }\href
  {https://doi.org/10.1103/PhysRevLett.94.230504} {\bibfield  {journal}
  {\bibinfo  {journal} {Physical Review Letters}\ }\textbf {\bibinfo {volume}
  {94}},\ \bibinfo {pages} {230504} (\bibinfo {year} {2005})}\BibitemShut
  {NoStop}%
\bibitem [{\citenamefont {Hwang}(2003)}]{Hwang.2003}%
  \BibitemOpen
  \bibfield  {author} {\bibinfo {author} {\bibfnamefont {W.-Y.}\ \bibnamefont
  {Hwang}},\ }\bibfield  {title} {\bibinfo {title} {Quantum key distribution
  with high loss: Toward global secure communication},\ }\href
  {https://doi.org/10.1103/PhysRevLett.91.057901} {\bibfield  {journal}
  {\bibinfo  {journal} {Physical Review Letters}\ }\textbf {\bibinfo {volume}
  {91}},\ \bibinfo {pages} {057901} (\bibinfo {year} {2003})}\BibitemShut
  {NoStop}%
\bibitem [{\citenamefont {Wang}(2005)}]{Wang.2005}%
  \BibitemOpen
  \bibfield  {author} {\bibinfo {author} {\bibfnamefont {X.-B.}\ \bibnamefont
  {Wang}},\ }\bibfield  {title} {\bibinfo {title} {Beating the
  photon-number-splitting attack in practical quantum cryptography},\ }\href
  {https://doi.org/10.1103/PhysRevLett.94.230503} {\bibfield  {journal}
  {\bibinfo  {journal} {Physical Review Letters}\ }\textbf {\bibinfo {volume}
  {94}},\ \bibinfo {pages} {230503} (\bibinfo {year} {2005})}\BibitemShut
  {NoStop}%
\bibitem [{\citenamefont {Lo}\ and\ \citenamefont {Preskill}(2007)}]{Lo.2007}%
  \BibitemOpen
  \bibfield  {author} {\bibinfo {author} {\bibfnamefont {H.-K.}\ \bibnamefont
  {Lo}}\ and\ \bibinfo {author} {\bibfnamefont {J.}~\bibnamefont {Preskill}},\
  }\bibfield  {title} {\bibinfo {title} {Security of quantum key distribution
  using weak coherent states with nonrandom phases},\ }\href@noop {} {\bibfield
   {journal} {\bibinfo  {journal} {Quantum Information {\&} Computation}\
  }\textbf {\bibinfo {volume} {8}},\ \bibinfo {pages} {431} (\bibinfo {year}
  {2007})}\BibitemShut {NoStop}%
\bibitem [{\citenamefont {Stucki}\ \emph {et~al.}(2005)\citenamefont {Stucki},
  \citenamefont {Brunner}, \citenamefont {Gisin}, \citenamefont {Scarani},\
  and\ \citenamefont {Zbinden}}]{Stucki.2005}%
  \BibitemOpen
  \bibfield  {author} {\bibinfo {author} {\bibfnamefont {D.}~\bibnamefont
  {Stucki}}, \bibinfo {author} {\bibfnamefont {N.}~\bibnamefont {Brunner}},
  \bibinfo {author} {\bibfnamefont {N.}~\bibnamefont {Gisin}}, \bibinfo
  {author} {\bibfnamefont {V.}~\bibnamefont {Scarani}},\ and\ \bibinfo {author}
  {\bibfnamefont {H.}~\bibnamefont {Zbinden}},\ }\bibfield  {title} {\bibinfo
  {title} {Fast and simple one-way quantum key distribution},\ }\href@noop {}
  {\bibfield  {journal} {\bibinfo  {journal} {Applied Physics Letters}\
  }\textbf {\bibinfo {volume} {87}},\ \bibinfo {pages} {194108} (\bibinfo
  {year} {2005})}\BibitemShut {NoStop}%
\bibitem [{\citenamefont {Ma}\ \emph {et~al.}(2005)\citenamefont {Ma},
  \citenamefont {Qi}, \citenamefont {Zhao},\ and\ \citenamefont
  {Lo}}]{Ma.2005}%
  \BibitemOpen
  \bibfield  {author} {\bibinfo {author} {\bibfnamefont {X.}~\bibnamefont
  {Ma}}, \bibinfo {author} {\bibfnamefont {B.}~\bibnamefont {Qi}}, \bibinfo
  {author} {\bibfnamefont {Y.}~\bibnamefont {Zhao}},\ and\ \bibinfo {author}
  {\bibfnamefont {H.-K.}\ \bibnamefont {Lo}},\ }\bibfield  {title} {\bibinfo
  {title} {Practical decoy state for quantum key distribution},\ }\href
  {https://doi.org/10.1103/PhysRevA.72.012326} {\bibfield  {journal} {\bibinfo
  {journal} {Physical Review A}\ }\textbf {\bibinfo {volume} {72}},\ \bibinfo
  {pages} {012326} (\bibinfo {year} {2005})}\BibitemShut {NoStop}%
\bibitem [{\citenamefont {Waks}\ \emph {et~al.}(2006)\citenamefont {Waks},
  \citenamefont {Takesue},\ and\ \citenamefont {Yamamoto}}]{Waks.2006}%
  \BibitemOpen
  \bibfield  {author} {\bibinfo {author} {\bibfnamefont {E.}~\bibnamefont
  {Waks}}, \bibinfo {author} {\bibfnamefont {H.}~\bibnamefont {Takesue}},\ and\
  \bibinfo {author} {\bibfnamefont {Y.}~\bibnamefont {Yamamoto}},\ }\bibfield
  {title} {\bibinfo {title} {Security of differential-phase-shift quantum key
  distribution against individual attacks},\ }\href
  {https://doi.org/10.1103/PhysRevA.73.012344} {\bibfield  {journal} {\bibinfo
  {journal} {Physical Review A}\ }\textbf {\bibinfo {volume} {73}},\ \bibinfo
  {pages} {012344} (\bibinfo {year} {2006})}\BibitemShut {NoStop}%
\bibitem [{\citenamefont {Stucki}\ \emph {et~al.}(2009)\citenamefont {Stucki},
  \citenamefont {Barreiro}, \citenamefont {Fasel}, \citenamefont {Gautier},
  \citenamefont {Gay}, \citenamefont {Gisin}, \citenamefont {Thew},
  \citenamefont {Thoma}, \citenamefont {Trinkler}, \citenamefont {Vannel} \emph
  {et~al.}}]{Stucki.2009}%
  \BibitemOpen
  \bibfield  {author} {\bibinfo {author} {\bibfnamefont {D.}~\bibnamefont
  {Stucki}}, \bibinfo {author} {\bibfnamefont {C.}~\bibnamefont {Barreiro}},
  \bibinfo {author} {\bibfnamefont {S.}~\bibnamefont {Fasel}}, \bibinfo
  {author} {\bibfnamefont {J.-D.}\ \bibnamefont {Gautier}}, \bibinfo {author}
  {\bibfnamefont {O.}~\bibnamefont {Gay}}, \bibinfo {author} {\bibfnamefont
  {N.}~\bibnamefont {Gisin}}, \bibinfo {author} {\bibfnamefont
  {R.}~\bibnamefont {Thew}}, \bibinfo {author} {\bibfnamefont {Y.}~\bibnamefont
  {Thoma}}, \bibinfo {author} {\bibfnamefont {P.}~\bibnamefont {Trinkler}},
  \bibinfo {author} {\bibfnamefont {F.}~\bibnamefont {Vannel}}, \emph
  {et~al.},\ }\bibfield  {title} {\bibinfo {title} {Continuous high speed
  coherent one-way quantum key distribution},\ }\href@noop {} {\bibfield
  {journal} {\bibinfo  {journal} {Optics Express}\ }\textbf {\bibinfo {volume}
  {17}},\ \bibinfo {pages} {13326} (\bibinfo {year} {2009})}\BibitemShut
  {NoStop}%
\bibitem [{\citenamefont {{Le Binh}}(2018)}]{LeBinh.2018}%
  \BibitemOpen
  \bibfield  {author} {\bibinfo {author} {\bibfnamefont {N.}~\bibnamefont {{Le
  Binh}}},\ }\href@noop {} {\emph {\bibinfo {title} {Optical modulation:
  Advanced techniques and applications in transmission systems and
  networks}}},\ Optical science and engineering\ (\bibinfo  {publisher}
  {{Taylor {\&} Francis a CRC title part of the Taylor {\&} Francis imprint a
  member of the Taylor {\&} Francis Group the academic division of T{\&}F
  Informa plc}},\ \bibinfo {address} {Boca Raton},\ \bibinfo {year}
  {2018})\BibitemShut {NoStop}%
\bibitem [{\citenamefont {Barnes}\ and\ \citenamefont
  {Barnes}(1993)}]{Barnes.1993}%
  \BibitemOpen
  \bibfield  {author} {\bibinfo {author} {\bibfnamefont {N.~P.}\ \bibnamefont
  {Barnes}}\ and\ \bibinfo {author} {\bibfnamefont {J.~C.}\ \bibnamefont
  {Barnes}},\ }\bibfield  {title} {\bibinfo {title} {Injection seeding i:
  Theory},\ }\href {https://doi.org/10.1109/3.250390} {\bibfield  {journal}
  {\bibinfo  {journal} {IEEE Journal of Quantum Electronics}\ }\textbf
  {\bibinfo {volume} {29}},\ \bibinfo {pages} {2670} (\bibinfo {year}
  {1993})}\BibitemShut {NoStop}%
\bibitem [{\citenamefont {Shirasaki}\ \emph {et~al.}(1988)\citenamefont
  {Shirasaki}, \citenamefont {Nishimoto}, \citenamefont {Okiyama},\ and\
  \citenamefont {Touge}}]{Shirasaki.1988}%
  \BibitemOpen
  \bibfield  {author} {\bibinfo {author} {\bibfnamefont {M.}~\bibnamefont
  {Shirasaki}}, \bibinfo {author} {\bibfnamefont {H.}~\bibnamefont
  {Nishimoto}}, \bibinfo {author} {\bibfnamefont {T.}~\bibnamefont {Okiyama}},\
  and\ \bibinfo {author} {\bibfnamefont {T.}~\bibnamefont {Touge}},\ }\bibfield
   {title} {\bibinfo {title} {Fibre transmission properties of optical pulses
  produced through direct phase modulation of dfb laser diode},\ }\href
  {https://doi.org/10.1049/el:19880330} {\bibfield  {journal} {\bibinfo
  {journal} {Electronics Letters}\ }\textbf {\bibinfo {volume} {24}},\ \bibinfo
  {pages} {486} (\bibinfo {year} {1988})}\BibitemShut {NoStop}%
\bibitem [{\citenamefont {{Chandra M. Natarajan}}\ \emph
  {et~al.}(2012)\citenamefont {{Chandra M. Natarajan}}, \citenamefont {{Michael
  G. Tanner}},\ and\ \citenamefont {{Robert H. Hadfield}}}]{Hadfield.2012}%
  \BibitemOpen
  \bibfield  {author} {\bibinfo {author} {\bibnamefont {{Chandra M.
  Natarajan}}}, \bibinfo {author} {\bibnamefont {{Michael G. Tanner}}},\ and\
  \bibinfo {author} {\bibnamefont {{Robert H. Hadfield}}},\ }\bibfield  {title}
  {\bibinfo {title} {Superconducting nanowire single-photon detectors: Physics
  and applications},\ }\href@noop {} {\bibfield  {journal} {\bibinfo  {journal}
  {Superconductor Science and Technology}\ }\textbf {\bibinfo {volume} {25}},\
  \bibinfo {pages} {063001} (\bibinfo {year} {2012})}\BibitemShut {NoStop}%
\bibitem [{\citenamefont {Jofre}\ \emph {et~al.}(2011)\citenamefont {Jofre},
  \citenamefont {Curty}, \citenamefont {Steinlechner}, \citenamefont {Anzolin},
  \citenamefont {Torres}, \citenamefont {Mitchell},\ and\ \citenamefont
  {Pruneri}}]{Jofre.2011}%
  \BibitemOpen
  \bibfield  {author} {\bibinfo {author} {\bibfnamefont {M.}~\bibnamefont
  {Jofre}}, \bibinfo {author} {\bibfnamefont {M.}~\bibnamefont {Curty}},
  \bibinfo {author} {\bibfnamefont {F.}~\bibnamefont {Steinlechner}}, \bibinfo
  {author} {\bibfnamefont {G.}~\bibnamefont {Anzolin}}, \bibinfo {author}
  {\bibfnamefont {J.~P.}\ \bibnamefont {Torres}}, \bibinfo {author}
  {\bibfnamefont {M.~W.}\ \bibnamefont {Mitchell}},\ and\ \bibinfo {author}
  {\bibfnamefont {V.}~\bibnamefont {Pruneri}},\ }\bibfield  {title} {\bibinfo
  {title} {True random numbers from amplified quantum vacuum},\ }\href
  {https://doi.org/10.1364/OE.19.020665} {\bibfield  {journal} {\bibinfo
  {journal} {Optics Express}\ }\textbf {\bibinfo {volume} {19}},\ \bibinfo
  {pages} {20665} (\bibinfo {year} {2011})}\BibitemShut {NoStop}%
\bibitem [{\citenamefont {Yuan}\ \emph {et~al.}(2014)\citenamefont {Yuan},
  \citenamefont {Lucamarini}, \citenamefont {Dynes}, \citenamefont
  {Fr{\"o}hlich}, \citenamefont {Plews},\ and\ \citenamefont
  {Shields}}]{Yuan.2014}%
  \BibitemOpen
  \bibfield  {author} {\bibinfo {author} {\bibfnamefont {Z.~L.}\ \bibnamefont
  {Yuan}}, \bibinfo {author} {\bibfnamefont {M.}~\bibnamefont {Lucamarini}},
  \bibinfo {author} {\bibfnamefont {J.~F.}\ \bibnamefont {Dynes}}, \bibinfo
  {author} {\bibfnamefont {B.}~\bibnamefont {Fr{\"o}hlich}}, \bibinfo {author}
  {\bibfnamefont {A.}~\bibnamefont {Plews}},\ and\ \bibinfo {author}
  {\bibfnamefont {A.~J.}\ \bibnamefont {Shields}},\ }\bibfield  {title}
  {\bibinfo {title} {Robust random number generation using steady-state
  emission of gain-switched laser diodes},\ }\href@noop {} {\bibfield
  {journal} {\bibinfo  {journal} {Applied Physics Letters}\ }\textbf {\bibinfo
  {volume} {104}},\ \bibinfo {pages} {261112} (\bibinfo {year}
  {2014})}\BibitemShut {NoStop}%
\end{thebibliography}%

\end{document}